\documentclass{v16nufact}

\def\be{\begin{equation}}
\def\ee{\end{equation}}
\def\bea{\begin{eqnarray}}
\def\eea{\end{eqnarray}}

\newcommand{\Photo}{}

\begin{document}
\vspace*{4cm}
\title{New Physics in Astrophysical Neutrino Flavor\\
NuFact 2016}

\author{Jordi Salvado$^1$, Carlos Arg\"uelles$^2$, Teppei Katori$^3$}
\address{$^1$Instituto de Fisica Corpuscular (IFIC), CSIC-Universitat de
  Valencia \\
  $^2$Massachusetts Institute of Technology (MIT) - Department of
  Physics\\
  $^3$Queen Mary University of London - School of Physics and Astronomy}

\maketitle

\abstract{Astrophysical neutrinos are powerful tools to study fundamental properties of particle
  physics. We perform a general new physics study on ultra
  high energy neutrino flavor content by introducing effective
  operators. We find that at the current limits on these operators,
  new physics terms cause maximal effects on the flavor content,
  however, the flavor content at Earth is confined to a region related
  to the initial flavor content.}

\section{Introduction}

In the last decades neutrino experiments have significantly improved
our knowledge of these elusive particles. All three angles of the
mixing matrix and the two mass differences have been measured. 
However, there are still a lot of questions to be answered. The
absolute value of the neutrino mass is still not known,
also the complex CP phase and the sign of one of the
mass differences are still unknown.

Most of the popular theoretical models that could naturally explain
such light masses propose a new mechanism that happens at higher
scales, thus introducing a new physics scale.  
Some of the models could also account for other unsolved question,
such as the origin of the observed baryon asymmetry in the Universe
\cite{Fukugita:1986hr,Shaposhnikov:2008pf,Hernandez:2016kel}, 
propose a Dark Matter candidate
\cite{Dodelson:1993je,Escudero:2016tzx}, or even explain the nature of Dark
Energy \cite{Simpson:2016gph}.  
In general we expect any new physics scale affect the
precision measurements of the oscillation phenomena for highly
energetic neutrinos or when observing neutrinos that traveled very
long distances \cite{GonzalezGarcia:2004wg,GonzalezGarcia:2005xw,Salvado:2016uqu}.

The last years the IceCube experiment discovered high energy
extraterrestrial neutrinos \cite{Aartsen:2013jdh}, 
probably coming from extra-galactic astrophysical sources, that 
have energies of order PeV and travel distances of order Mpc-Gpc.
The IceCube detector is located in the South Pole and is a
kilometer cube of instrumented ice situated more than a kilometer
deep in the Antarctic ice. The ice is instrumented with 86
strings of 60 digital optical modules each one has a photo-multiplier for collecting photons.
Due to the deep isolation photons are mainly Cerenkov
light of charged particles traveling through the ice.

In 4 years 54 neutrinos like events with energies above 20TeV where observe in the
IceCube detector at the South Pole \cite{Aartsen:2014gkd}. The High Energy Starting Events
analysis (HESE), due to the veto region, allows
the use of the full sky.
There are two main topologies that
can be distinguished: the tracks, produced by muons crossing the
detector, and the showers, that can be produce by any neutrino flavor
via neutral current interaction or by electron/tau neutrino charged current
interactions.
The track events are mainly produced by the muon neutrinos.
The different topologies may give us information about the flavor
content of the astrophysical neutrino flux. However, extracting the information
is complicated since there are non trivial
correlations with the amount of neutrinos and anti-neutrinos, and also with
the spectral features of the astrophysical neutrino flux \cite{Mena:2014sja,Aartsen:2015ivb,Vincent:2016nut}. 
More statistics is needed to extract more precise information
about the flavor ratio of the astrophysical
neutrinos, which may be achieved in the next generation experiments
Icecube-Gen2 \cite{Aartsen:2014njl} and KM3NeT \cite{Adrian-Martinez:2016fdl}.

Measuring the flavor ratio of these high energetic neutrinos could
have implications on neutrino physics \cite{Arguelles:2015dca,Bustamante:2015waa,Katori:2016eni}. 

\section{New Physics in Neutrino Oscillations}
Neutrino oscillations are quantum effects related to the 
misalignment between mass and flavor eigenstates.
In vacuum the
neutrino propagation is well described by a three
level quantum system where the Hamiltonian is given by
\begin{equation}
  H=\frac{1}{2E}U M^2 U^\dagger,
\end{equation}
with $U$ being a unitary matrix that relates the propagation and flavor
eigenstates space(in vacuum the propagation eigenstates are also called mass
eigenstates), and
$M$ is a  diagonal matrix that contains the mass eigenvalues.
The mixing matrix $U$ and the mass square differences in $M^2$ are
measured by neutrino oscillation experiments \cite{Gonzalez-Garcia:2014bfa}.

In the case of neutrinos propagating very large distances  the wave
packages of the different mass eigenstates will not overlap and the
oscillation phenomena stops, in this case the probability of measuring
a flavor $\alpha$ for a neutrino produced as a flavor $\beta$ is given by,

\begin{equation}
  \label{prob}
  \bar P_{\nu_{\alpha} \to \nu_{\beta}} =
  \sum_{i}\left|U_{\alpha i}\right|^2\left|U_{\beta
      i}\right|^2~.
\end{equation}

The decoherence process of neutrinos and the size of the wave packages is
still unknown, but for astrophysical distances this is a very good approximation.

In the presence of new physics neutrino oscillations are modified by a
new operator in the propagating Hamiltonian, 

\begin{equation}
  H=\frac{1}{2E}U M^2 U^\dagger +{\sum_n \left(\frac{E}{\Lambda_n}\right)^n \tilde U_n O_n \tilde U_n^\dagger}
  = V^\dagger(E) \Delta V(E),
\end{equation}

where $O_n$ and $\Lambda_n$ define the scale of the new physics. 
The bounds in this operators are well studied in different context. In
particular, for the power $O_0$ and $O_1$ are shown
in Table \ref{table_bounds}. 
The matrix that relates the
propagation and flavor eigenstates is not $U$, but the unitary matrix
$V(E)$ that depends on the neutrino energy.
Notice that since the new physics is expected to be relevant at
energies of order of the cutoff $\Lambda_n$ or higher the IceCube
neutrinos with the highest energies ever measuerd, could give relevant information about the new physics.

\begin{table}

\begin{center}
  \begin{tabular}{|c|l|l|}
    \hline\hline 
    $n$ & New Physics & Current Bound\\ 
    &  & {\small From SK \cite{Abe:2014wla}   and IC-atm \cite{Abbasi:2010kx}}\\ 
    \hline 
    0 & CPT-odd Lorentz Violation & $O_0<10^{-23}$GeV \\ 
    & Coupling Space Time Torsion &\\
    &  Non Standard Neutrino Interactions & \\
    \hline 
    1 & CPT-even Lorentz Violation & $O_1/\Lambda_1 < 10^{-27}$\\
    & Violation of the equivalence principle & \\
    \hline 
  \end{tabular} \\
\caption{\label{table_bounds} Current bounds from Super Kamiokande
   and IceCube-atmospheric for
  new physics scales.}
  \end{center}
\end{table}

The oscillation probability formula is the same as eq. (\ref{prob})
just replacing the matrix $U$ by the matrix $V(E)$, notice the energy dependence,
\begin{equation}
  \label{probnew}
  \bar P_{\nu_{\alpha} \to \nu_{\beta}}(E) =
  \sum_{i}\left|V_{\alpha i}(E)\right|^2\left|V_{\beta
      i}(E)\right|^2~.
\end{equation}

\section{Flavor Ratio at Detection}

In order to show the result we plot probabilities densities of
measuring a flavor ratio at detection point assuming some flavor ratio at the
production.
We show the result for different values of the new physics scales
using a Bayesian approach in order to have into account the unknown of
the flavor structure of the new physics we assume a flat distribution
in the Haar measure of $SU(3)$ for
the matrices $\tilde U$,
\begin{equation}
  d\tilde U_n = d\tilde s^2_{12} \wedge d\tilde c^4_{13} \wedge d\tilde s^2_{23} \wedge d\tilde\delta~,
\end{equation}
For the uncertainty on the current oscillation parameters we use the results in www.nu-fit.org \cite{Gonzalez-Garcia:2014bfa}.

In the following we will show the results for the different possible
production fluxes:
\begin{itemize}
\item Charged Pion Decay $(1:2:0)$. Pions are produced via a 
  $\Delta$ resonance and the following decays produce 1 electron and
  2 muon neutrinos.

\item Pion Decay and Muon Energy Lost $(0:1:0)$. Like the pion
  decay scenario but in this case due to synchrotron cooling the secondary
  Muon loses part of the energy before decaying, making the resulting
  neutrinos effectively disappear from the high energy IceCube
  events.

\item Neutron Decay $(1:0:0)$, pure astrophysical neutron source.
  
\item Production of $\nu_\tau$ $(0:0:1)$. We add this by
  completeness, it's not motivated by any known physical scenario but
  measuring this may very good be a hint of some new physics
  production mechanism. 
\end{itemize}

\begin{figure}[ht]
  \begin{center}
    \includegraphics[width=10cm]{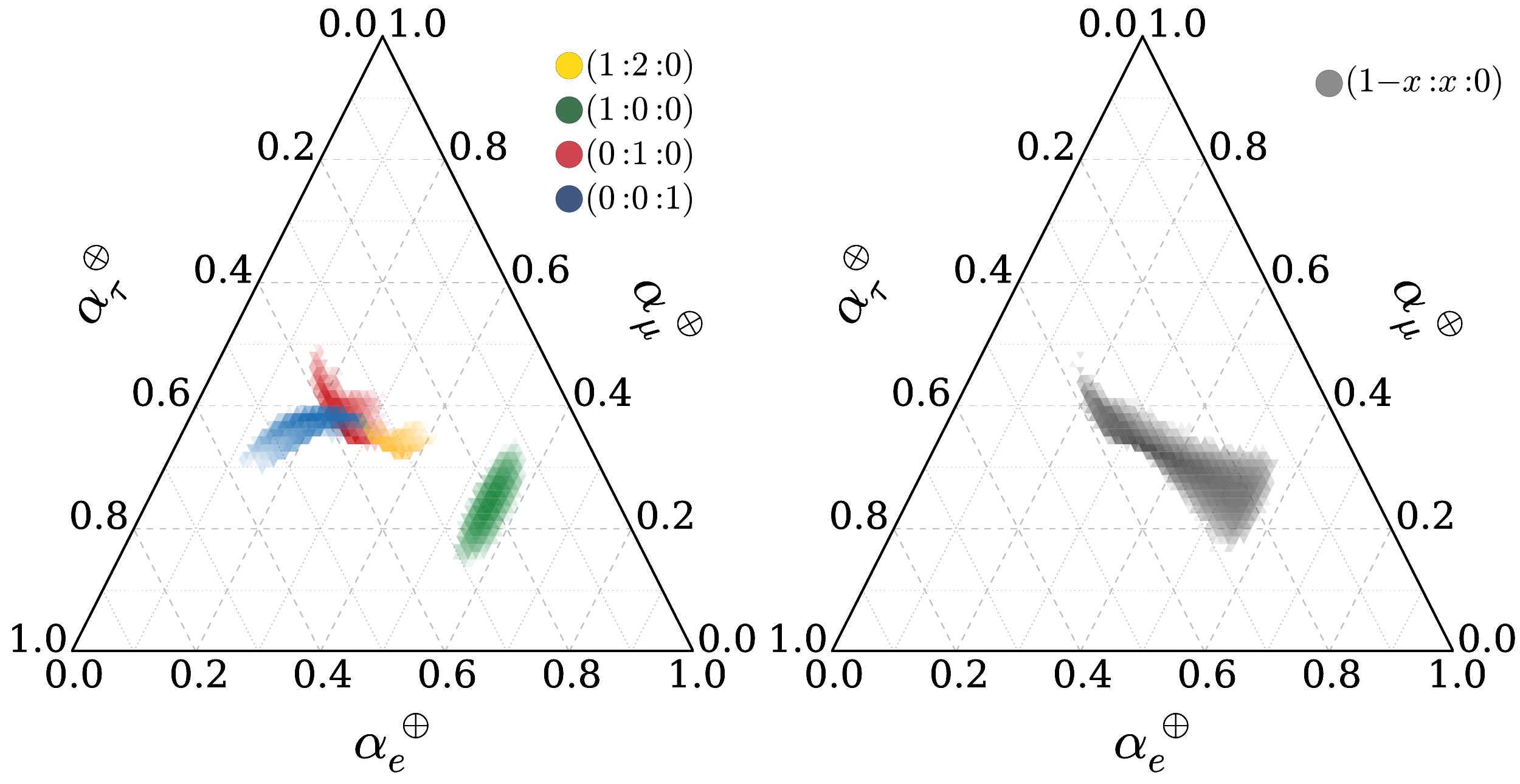}
    \caption{ \label{fig:nonewpy}Non new physics case, here we only show the effect of the
      current uncertainties in the mixing parameters from neutrino
      oscillation experiments. In the right we show the case for the 4
      benchmark cases described above and in the right the result of a
      case where both of them are combined with an unknown relative weighting.}
  \end{center}

\end{figure}

In Fig.\ref{fig:nonewpy} we show the result only with the current
uncertainties in the oscillation parameters. Notice that the current
precision still give some freedom but independently of the initial
flux the allowed parameter region is relatively small.

\begin{figure}[ht]
  \begin{center}
    \includegraphics[width=10cm]{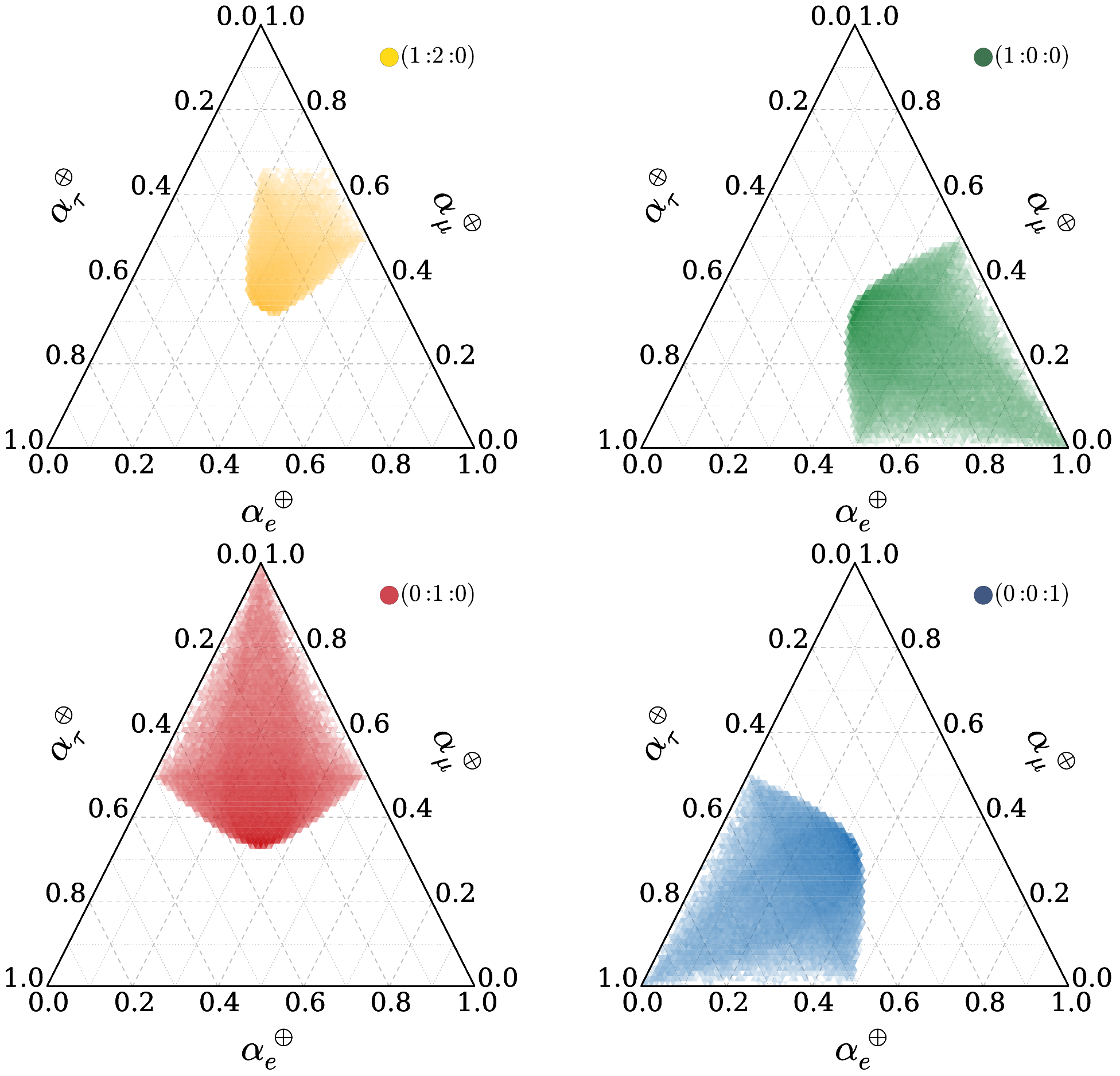}
    \caption{\label{fig:onlynewpy}Only new physics case, here we only show the effect for
      totally new physics dominated case.}
  \end{center}
    
\end{figure}

In Fig.\ref{fig:onlynewpy} we show the case totally dominated by the new
physics scenario, in here the mass term in the Hamiltonian is
neglected and we can see the maximum effect for the new physics.
In this case,we can see how even in this case the initial information
of the flavor at production is still pretty conserve since the overlap
of the regions is marginal.

\begin{figure}[t]
  \begin{center}
    \includegraphics[width=9cm]{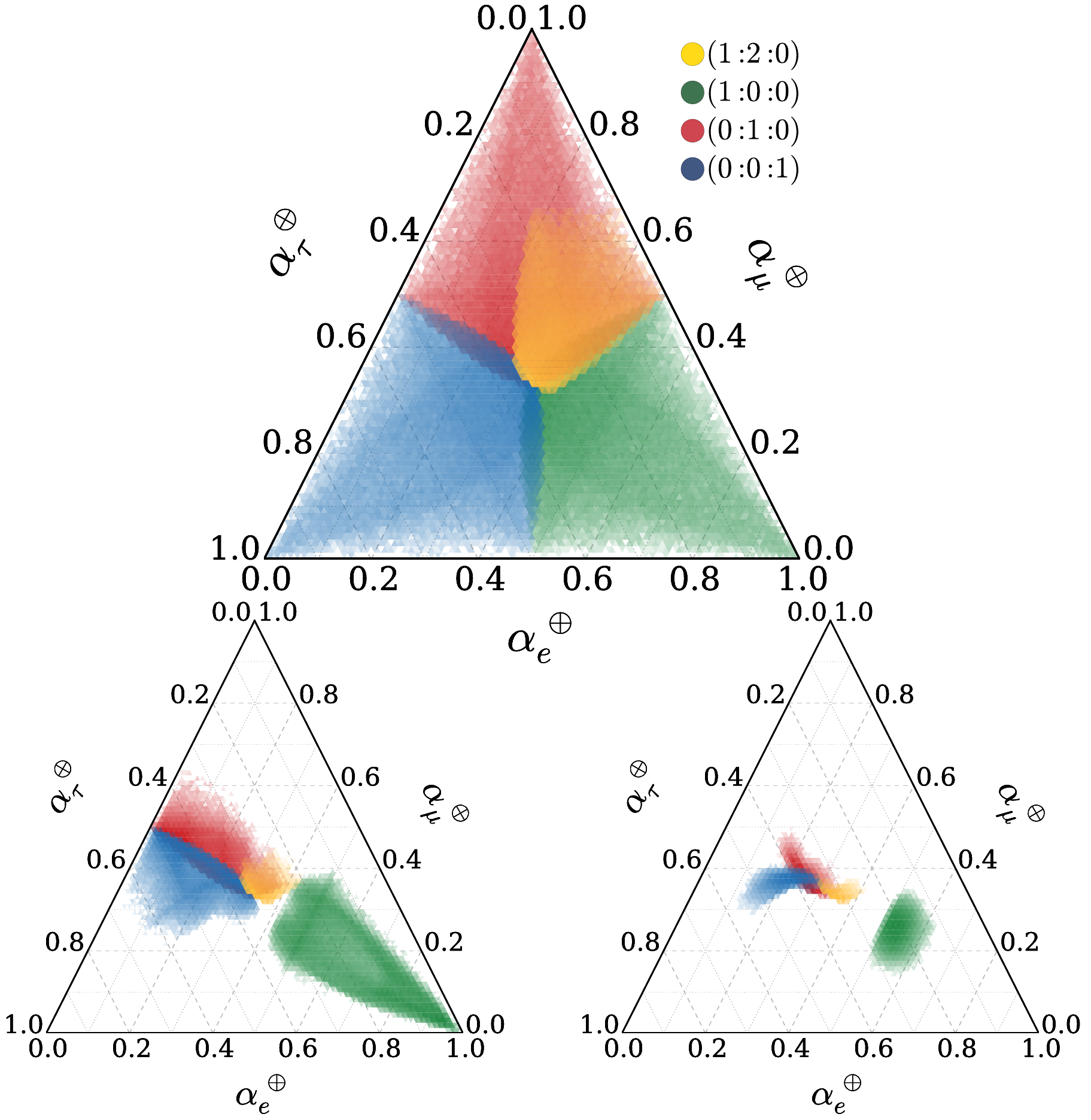}
    \caption{ \label{new0}Case $n=0$ with three different values of the new physics scale,
      $O_0=10^{-23}$ top, $O_0=3.6\times 10^{-26}$ bottom left, and
      $O_0=6.3\times 10^{-28}$ bottom right.}
  \end{center}
\end{figure}

\begin{figure}[t]
  \begin{center}
    \includegraphics[width=9cm]{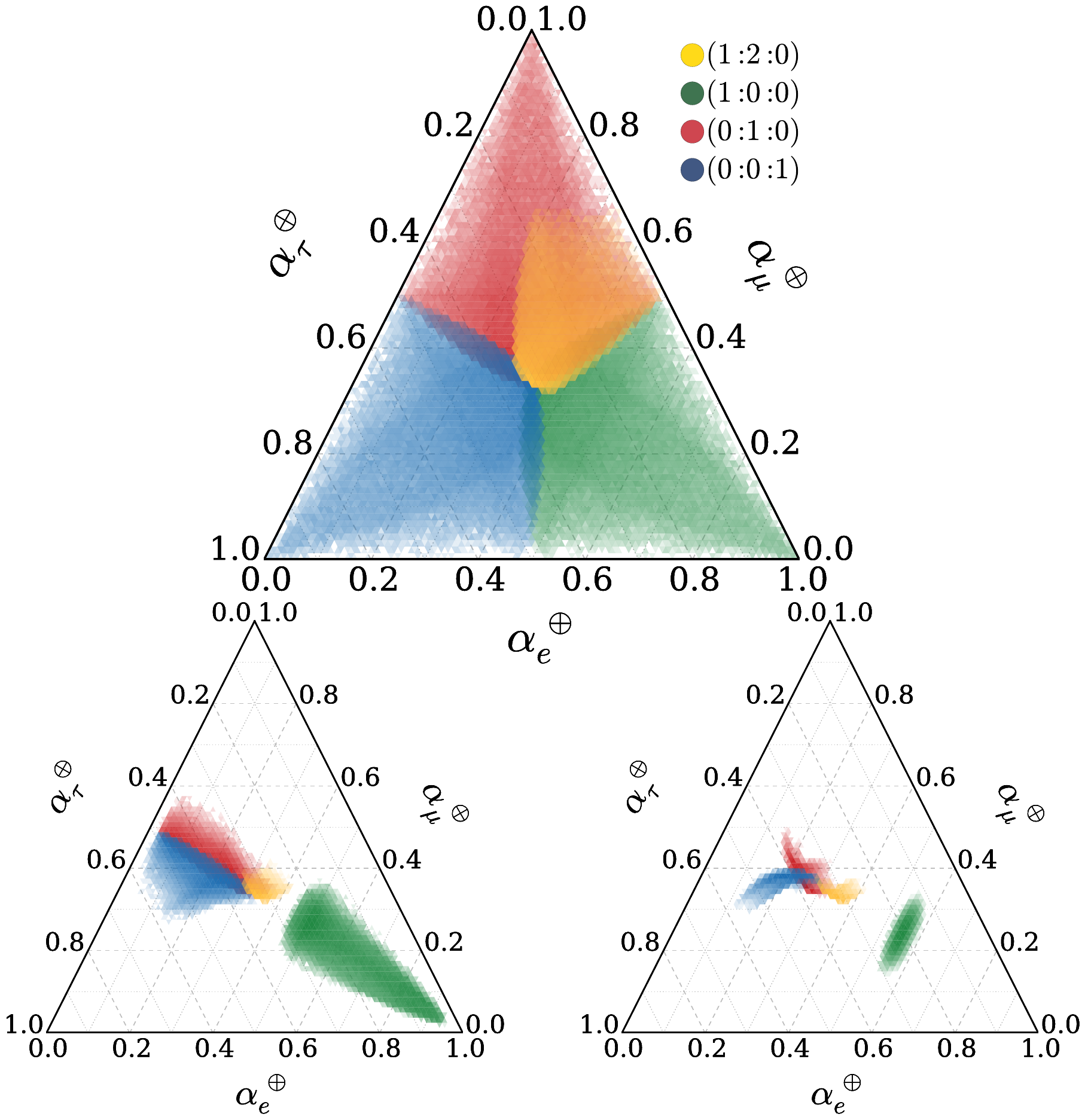}
    \caption{  \label{new01}Case $n=1$ with three different values of the new physics scale,
      $O_1/\Lambda_1=10^{-27}$ top, $O_1/\Lambda_1=1.0\times 10^{-30}$ bottom left, and
      $O_1/\Lambda_1=3.2\times 10^{-34}$ bottom right.}
  \end{center}
\end{figure}

In the following we will see the intermediate case where both terms
are relevant for a different value of the the new physics scale, in the
case $n=0$ and $n=1$ this is shown in figures Fig.\ref{new0} and
Fig.\ref{new01}.


Notice that in both cases $n=0$ and $n=1$ the bottom plot is done
with a new physics scale of the same magnitude of the current bounds
which directly implies that any measure would put a the strongest
bound.

\section{Conclusions}

Fig.\ref{new0} and Fig.\ref{new01} show how sensitive is the
astrophysical neutrino flavor ratio to the new physics and how this
relatively decouple from the initial productions fluxes. 
We can reach the following conclusions:
\begin{itemize}
\item The measurement of the flavor content at any precision
  would give the strongest bounds or discovery potential of the new
  physics scenarios. 
\item The regions do not overlap significantly, therefore, some
  flavor information in the production is always preserved. 
  A consequence of this would be that a measurement of tau neutrinos
  dominated flux outside the standard oscillation regions 
  will imply new physics in the production and propagation.
\item Sizable effect in the measure is given by new physics scales
  that are three orders of magnitude smaller than the current bounds. 
\item The conclusion is robust for both cases $n=0$ and $n=1$, larger
  $n$ does not change significantly the result. 
\item If the production is via pion decay the new physics is confined in a small region.

\end{itemize}

\clearpage
\section*{Acknowledgments}

This work was partially supported by grants FPA2014-57816-P,
PROMETEOII/2014/050, and  the European projects
H2020-MSCA-ITN-2015//674896-ELUSIVES and H2020-MSCA-RISE-2015.


\section*{References}

\begin{thebibliography}{99}

  \bibitem{Fukugita:1986hr}
  M.~Fukugita and T.~Yanagida,
  Phys.\ Lett.\ B {\bf 174}, 45 (1986).
  doi:10.1016/0370-2693(86)91126-3

  \bibitem{Shaposhnikov:2008pf}
  M.~Shaposhnikov,
  JHEP {\bf 0808}, 008 (2008)
  doi:10.1088/1126-6708/2008/08/008
  [arXiv:0804.4542 [hep-ph]].
\bibitem{Hernandez:2016kel}
  P.~Hernández, M.~Kekic, J.~López-Pavón, J.~Racker and J.~Salvado,
  JHEP {\bf 1608}, 157 (2016)
  doi:10.1007/JHEP08(2016)157
  [arXiv:1606.06719 [hep-ph]].

  \bibitem{Dodelson:1993je}
  S.~Dodelson and L.~M.~Widrow,
  Phys.\ Rev.\ Lett.\  {\bf 72}, 17 (1994)
  doi:10.1103/PhysRevLett.72.17
  [hep-ph/9303287].

\bibitem{Escudero:2016tzx}
  M.~Escudero, N.~Rius and V.~Sanz,
  arXiv:1606.01258 [hep-ph].

  
\bibitem{Simpson:2016gph}
  F.~Simpson, R.~Jimenez, C.~Pena-Garay and L.~Verde,
  arXiv:1607.02515 [astro-ph.CO].



  \bibitem{GonzalezGarcia:2004wg}
  M.~C.~Gonzalez-Garcia and M.~Maltoni,
  Phys.\ Rev.\ D {\bf 70} (2004) 033010
  doi:10.1103/PhysRevD.70.033010
  [hep-ph/0404085].

\bibitem{GonzalezGarcia:2005xw}
  M.~C.~Gonzalez-Garcia, F.~Halzen and M.~Maltoni,
  Phys.\ Rev.\ D {\bf 71} (2005) 093010
  doi:10.1103/PhysRevD.71.093010
  [hep-ph/0502223].

\bibitem{Bustamante:2015waa}
  M.~Bustamante, J.~F.~Beacom and W.~Winter,
  Phys.\ Rev.\ Lett.\  {\bf 115} (2015) no.16,  161302
  doi:10.1103/PhysRevLett.115.161302
  [arXiv:1506.02645 [astro-ph.HE]].
  
\bibitem{Salvado:2016uqu}
  J.~Salvado, O.~Mena, S.~Palomares-Ruiz and N.~Rius,
  arXiv:1609.03450 [hep-ph].
  
\bibitem{Aartsen:2013jdh}
  M.~G.~Aartsen {\it et al.} [IceCube Collaboration],
  Science {\bf 342} (2013) 1242856
  doi:10.1126/science.1242856
  [arXiv:1311.5238 [astro-ph.HE]].
\bibitem{Aartsen:2014gkd}
  M.~G.~Aartsen {\it et al.} [IceCube Collaboration],
  Phys.\ Rev.\ Lett.\  {\bf 113} (2014) 101101
  doi:10.1103/PhysRevLett.113.101101
  [arXiv:1405.5303 [astro-ph.HE]].


\bibitem{Mena:2014sja}
  O.~Mena, S.~Palomares-Ruiz and A.~C.~Vincent,
  Phys.\ Rev.\ Lett.\  {\bf 113} (2014) 091103
  doi:10.1103/PhysRevLett.113.091103
  [arXiv:1404.0017 [astro-ph.HE]].

  \bibitem{Aartsen:2015ivb}
  M.~G.~Aartsen {\it et al.} [IceCube Collaboration],
  Phys.\ Rev.\ Lett.\  {\bf 114} (2015) no.17,  171102
  doi:10.1103/PhysRevLett.114.171102
  [arXiv:1502.03376 [astro-ph.HE]].

  
  \bibitem{Vincent:2016nut}
  A.~C.~Vincent, S.~Palomares-Ruiz and O.~Mena,
  Phys.\ Rev.\ D {\bf 94} (2016) no.2,  023009
  doi:10.1103/PhysRevD.94.023009
  [arXiv:1605.01556 [astro-ph.HE]].

\bibitem{Aartsen:2014njl}
  M.~G.~Aartsen {\it et al.} [IceCube Collaboration],
  arXiv:1412.5106 [astro-ph.HE].

\bibitem{Adrian-Martinez:2016fdl}
  S.~Adrian-Martinez {\it et al.} [KM3Net Collaboration],
  J.\ Phys.\ G {\bf 43}, no. 8, 084001 (2016)
  doi:10.1088/0954-3899/43/8/084001
  [arXiv:1601.07459 [astro-ph.IM]].
  
\bibitem{Arguelles:2015dca}
  C.~A.~Argüelles, T.~Katori and J.~Salvado,
  Phys.\ Rev.\ Lett.\  {\bf 115} (2015) 161303
  doi:10.1103/PhysRevLett.115.161303
  [arXiv:1506.02043 [hep-ph]].

\bibitem{Katori:2016eni}
  T.~Katori, C.~A.~Argüelles and J.~Salvado,
  arXiv:1607.08448 [hep-ph].
  
  \bibitem{Gonzalez-Garcia:2014bfa}
  M.~C.~Gonzalez-Garcia, M.~Maltoni and T.~Schwetz,
  JHEP {\bf 1411} (2014) 052
  doi:10.1007/JHEP11(2014)052
  [arXiv:1409.5439 [hep-ph]].


\bibitem{Abe:2014wla}
  K.~Abe {\it et al.} [Super-Kamiokande Collaboration],
  Phys.\ Rev.\ D {\bf 91}, no. 5, 052003 (2015)
  doi:10.1103/PhysRevD.91.052003
  [arXiv:1410.4267 [hep-ex]].
\bibitem{Abbasi:2010kx}
  R.~Abbasi {\it et al.} [IceCube Collaboration],
  Phys.\ Rev.\ D {\bf 82}, 112003 (2010)
  doi:10.1103/PhysRevD.82.112003
  [arXiv:1010.4096 [astro-ph.HE]].



\end{thebibliography}
\end{document}